%% file: paper.tex
\pdfoutput=1

\documentclass[11pt]{article}

\usepackage[final]{acl}

\usepackage{times}
\usepackage{latexsym}

\usepackage[T1]{fontenc}

\usepackage[utf8]{inputenc}

\usepackage{microtype}

\usepackage{inconsolata}

\usepackage{hyperref}
\usepackage{url}
\usepackage{xspace}
\usepackage{booktabs}
\usepackage{graphicx}
\usepackage{algorithm}
\usepackage{multirow}
\usepackage{algorithmic}
\usepackage{pgfplots}
\usepackage{makecell}
\usepackage{listings}
\usepackage{titlesec}
\usepackage{amsmath}
\usepackage{pifont}
\usepackage{arydshln}
\usepackage{tcolorbox}
\tcbuselibrary{breakable}
\usepackage[fencedCode]{markdown}
\usepackage{enumitem}
\usepackage{makecell}
\usepackage{cleveref}
\usepackage{xcolor}

\crefname{section}{§}{§§}
\Crefname{section}{§}{§§}

\definecolor{DarkGreen}{RGB}{30,130,30}
\newcommand{\cmark}{\textcolor{DarkGreen}{\ding{51}}}
\newcommand{\xmark}{\textcolor{red}{\ding{55}}}%

\newcommand{\model}{{OpenAgent}\xspace}
\newcommand{\bench}{{OpenAct}\xspace}
\newcommand{\email}[1]{\href{mailto:#1}{\nolinkurl{#1}}}

\title{Enhancing Open-Domain Task-Solving Capability of LLMs \\ via Autonomous Tool Integration from GitHub}

\author{
    \textbf{Bohan Lyu\textsuperscript{1}\footnotemark[1]},
    \textbf{Xin Cong\textsuperscript{1}\footnotemark[1]\footnotemark[2]},
    \textbf{Heyang Yu\textsuperscript{1}},
    \textbf{Pan Yang\textsuperscript{1}}, 
    \textbf{Cheng Qian\textsuperscript{1,3}},
    \textbf{Zihe Wang\textsuperscript{1}}, \\
    \textbf{Yujia Qin\textsuperscript{1}}, 
    \textbf{Yining Ye\textsuperscript{1}},
    \textbf{Yaxi Lu\textsuperscript{1}},
    \textbf{Chen Qian\textsuperscript{1,4}},
    \textbf{Zhong Zhang\textsuperscript{1}}, 
    \textbf{Yukun Yan\textsuperscript{1}},\\
    \textbf{Yankai Lin\textsuperscript{2}},
    \textbf{Zhiyuan Liu\textsuperscript{1}\footnotemark[2]}, 
    \textbf{Maosong Sun\textsuperscript{1}}
    \\
\textbf{\textsuperscript{1}} Department of Computer Science and Technology, Tsinghua University \\
\textbf{\textsuperscript{2}} Gaoling School of Artificial Intelligence, Renmin University of China \\
\textbf{\textsuperscript{3}} University of Illinois Urbana-Champaign \\
\textbf{\textsuperscript{4}} School of Artificial Intelligence, Shanghai Jiao Tong University \\
\texttt{\small lvbh22@mails.tsinghua.edu.cn,congxin1995@tsinghua.edu.cn} 
}

\begin{document}
\maketitle

\renewcommand{\thefootnote}{\fnsymbol{footnote}}
\footnotetext[1]{\ \ Equal contribution.}
\footnotetext[2]{\ \ Corresponding author.}
\renewcommand{\thefootnote}{\arabic{footnote}}

\begin{abstract}
Large Language Models (LLMs) excel in traditional natural language processing tasks but struggle with problems that require complex domain-specific calculations or simulations. While equipping LLMs with external tools to build LLM-based agents can enhance their capabilities, existing approaches lack the flexibility to address diverse and ever-evolving user queries in open domains. Currently, there is also no existing dataset that evaluates LLMs on open-domain knowledge that requires tools to solve.
%
%
To this end, we introduce \textbf{\bench} benchmark to evaluate the open-domain task-solving capability, which is built on human expert consultation and repositories in GitHub. It comprises 339 questions spanning 7 diverse domains that need to be solved with domain-specific methods. In our experiments, even state-of-the-art LLMs and LLM-based agents demonstrate unsatisfactory success rates, underscoring the need for a novel approach.
Furthermore, we present \textbf{\model}, a novel LLM-based agent system that can tackle evolving queries in open domains through autonomously integrating specialized tools from GitHub. \model employs 1) a hierarchical framework where specialized agents handle specific tasks and can assign tasks to inferior agents, 2) a bi-level experience learning mechanism to learn from both humans' and its own experiences to tackle tool flaws. Experiments demonstrate its superior effectiveness and efficiency, which significantly outperforms baselines. Our data and code are open-source at \url{https://github.com/OpenBMB/OpenAct}.
\end{abstract}

\input{section/intro.tex}
\input{section/related_work.tex}

\input{section/dataset}
\input{section/method.tex}
\input{section/exp.tex}
\input{section/conclusion.tex}
\input{section/limitation.tex}

\bibliography{custom,ref}

\appendix

\input{section/appendix.tex}

\end{document}

%% file: section/intro.tex
\section{Introduction}

\begin{figure}[!t]
    \centering
    \includegraphics[width=\linewidth]{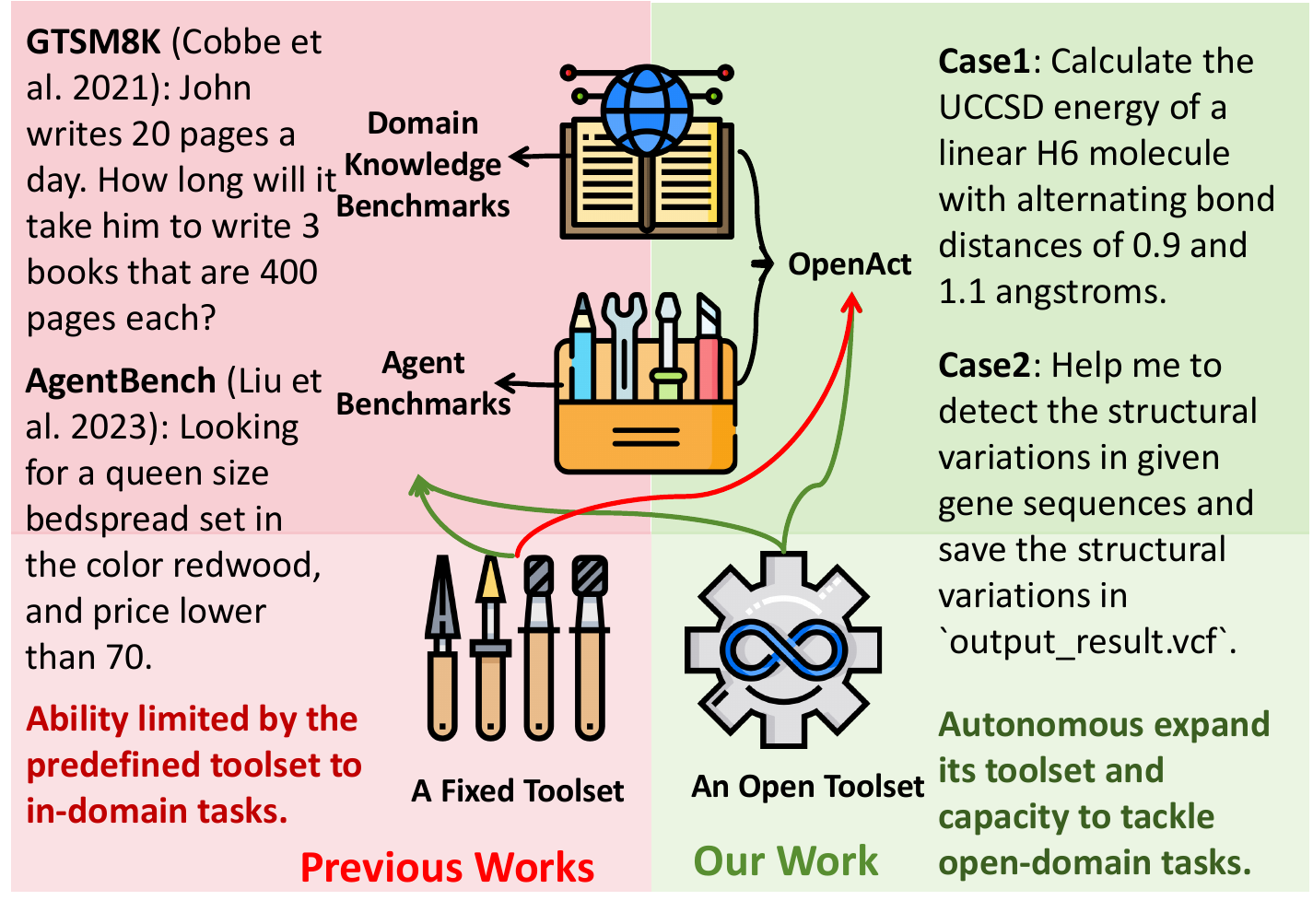}
    \caption{The comparison between our work and previous studies from both dataset and model perspectives. \begin{minipage}[b]{0.04\columnwidth} \centering \includegraphics[width=\linewidth]{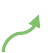} \end{minipage} indicates that a model is capable of solving tasks within a benchmark, whereas \begin{minipage}[b]{0.04\columnwidth} \centering \includegraphics[width=\linewidth]{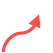} \end{minipage} indicates the opposite.}
    \label{fig:case1}
\end{figure}

Large Language Models (LLMs) have demonstrated exceptional capabilities through diverse traditional natural language processing (NLP) tasks~\citep{openaichatgptblog, openai2023gpt4, team2023gemini, anthropic2024claude3}.
However, LLMs still struggle with specialized tasks that require calculation, simulation, data augmentation, etc~\citep{qin2024toollearningfoundationmodels}.
To tackle it, researchers equip LLMs with external tools (e.g., search engines~\citep{nakano2021webgpt,qin2023webcpm}, code executors~\citep{qian2023creator, cai2023large}, scientific simulators~\citep{liu2022mindseyegroundedlanguage, bran2023chemcrowaugmentinglargelanguagemodels}) to function as agents that are capable of solving complex tasks and extend the capability boundary of LLMs beyond traditional NLP tasks.
%
%
Existing LLM-based agents~\citep{autogpt, xagent2023, schick2023toolformer, parisi2022talm, patil2023gorilla, hong2024metagptmetaprogrammingmultiagent, lyu2024adapting} have access to a pre-defined toolset and can combine their cognitive abilities with the specialized functionalities of these tools.

However, the effectiveness of current LLM agents is constrained by the pre-defined toolset.
This paradigm restricts them to addressing in-domain problems while lacking the generalization capability to handle diverse open-domain questions, which is a critical limitation given that real-world user queries often require domain-specialized tools.
This contradiction raises our core research problem: \textbf{Are LLM agents able to autonomously search for and adapt new tools for open-domain tasks?}

Addressing this challenge requires rethinking existing evaluation paradigms. Current benchmarks are constructed based on pre-defined toolsets rather than real-world demands, creating an artificial performance ceiling that fails to assess models' ability to handle tasks requiring external specialized tools.
Real-world tasks often necessitate specialized tools and domain-specific knowledge that extend beyond the inherent capabilities of pre-trained language models and a fixed toolset. Such tasks, like gene mutation detection, quantum chemistry analysis, and financial modeling, are typically executed by domain experts utilizing sophisticated professional tools and software.

In this context, GitHub emerges as it is the largest platform that contains implementations of algorithms and methodologies in open domains that are employed by experts in their respective fields. 
%
%
If LLM-based agents could effectively search for, deploy, and utilize relevant repositories from GitHub, they could autonomously extend their tool set. This capability would enable LLM agents to dynamically adapt and grow their abilities, significantly enhancing their versatility and effectiveness in addressing real-world open-domain tasks.

\begin{figure}[!t]
    \centering
    \includegraphics[width=\linewidth]{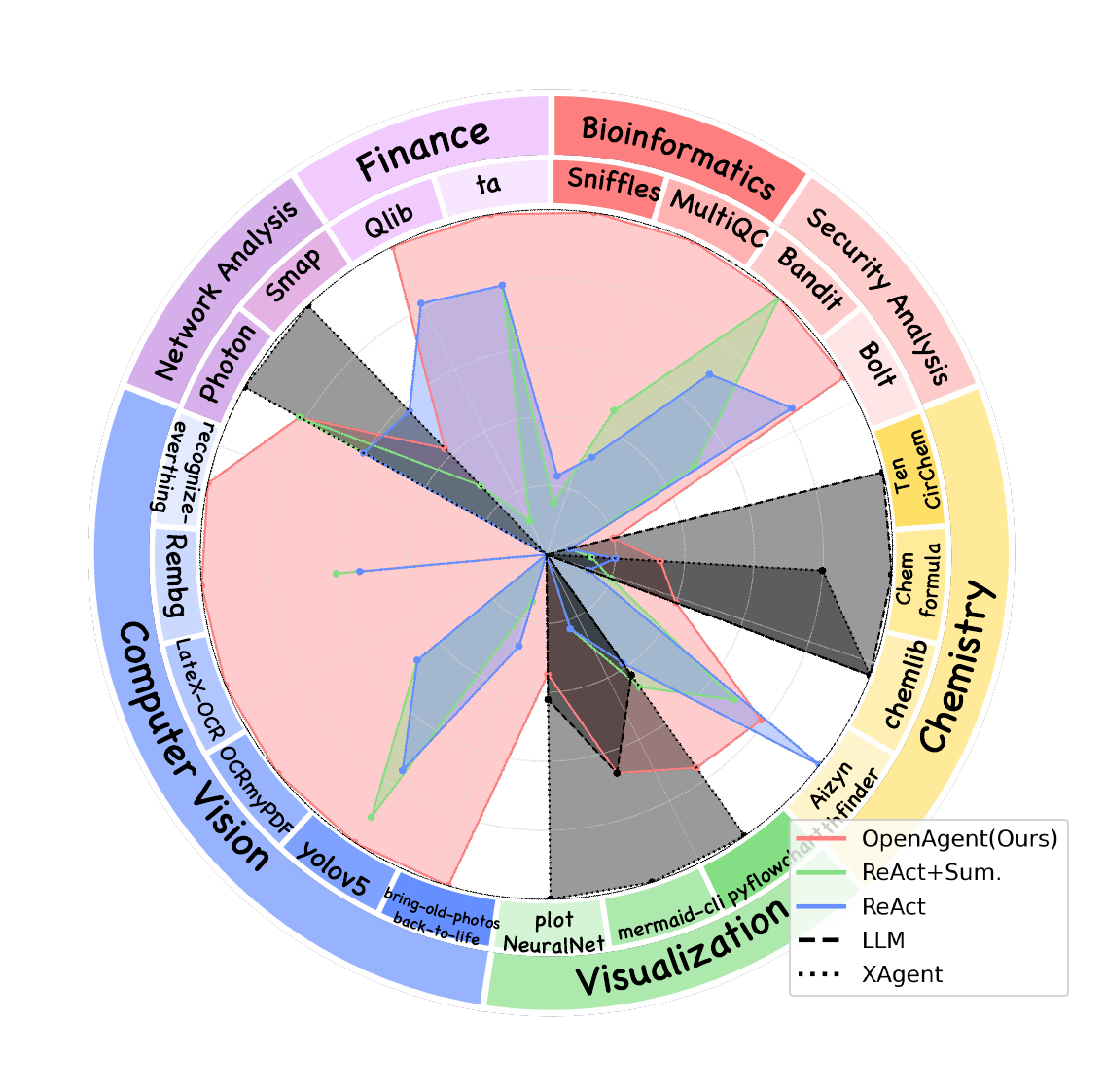}
    \caption{Illustration of GPT-4-based \model performs against baselines on $21$ repositories of $7$ domains in \bench.}
    \label{fig:intro-pie}
\end{figure}

To this end, we introduce a novel benchmark \textbf{\bench}. Its construction began by identifying key issues and methodologies across multiple domains. We then collected relevant tools on GitHub and carefully designed a series of tasks that reflect domain needs. It comprises 339 queries across 7 diverse domains, including finance, chemistry, bioinformatics, computer vision, etc. Compared with existing benchmarks, \bench is the first dataset designed to evaluate LLMs on fulfilling open-domain tasks (see Figure~\ref{fig:case1}).
%
%
Experimental results show that both vanilla LLMs and general-purpose LLM agents perform poorly on \bench (see Figure~\ref{fig:intro-pie}).
The key challenges are: (1) Lack of Quality Assurance: GitHub repositories often contain flaws, bugs, or incomplete/misleading documentation, (2) Alignment Gap: GitHub tools require adjustments to fit user needs, (3) Workflow Complexity: The process involves many diverse tasks, making it hard for LLMs to stay effective. 

To solve these challenges, we introduce \textbf{\model}, a novel LLM-based agent system that autonomously extends tools from GitHub. It possesses two key features: (1) a hierarchical structure that dynamically decomposes the whole process into distinct subtasks~(e.g., setting up environments, reading Issues/PRs), and (2) a bi-level experience learning mechanism to accumulate both in-task and cross-task experiences. Generally, \model starts with searching suitable repositories, then setting up the necessary environment, and utilizing the repository to fulfill user queries.

We conduct extensive experiments on \bench to demonstrate the effectiveness of \model compared to state-of-the-art LLMs and LLM agents. Ablation studies prove the necessity of both of the above two key features.

In summary, our contributions are threefold:
\begin{itemize}
    \item We introduce \bench, a comprehensive dataset comprising 339 queries across 7 diverse domains, which is specifically designed to evaluate the capabilities of open-domain task-solving capability of LLMs.
    \item We propose \model, a novel LLM agent system that autonomously extends its tool set by integrating repositories from GitHub. It employs a hierarchical structure and possesses a bi-level experience learning mechanism.
    \item Experiments demonstrate the effectiveness of \model against baselines on \bench, and extensive experiments also validate our proposed two key features in \model.
\end{itemize}

%% file: section/related_work.tex
\begin{table*}[t]
\centering
\resizebox{\textwidth}{!}{
\begin{tabular}{c|ccccccc}
\toprule
Benchmark & Domain Num.& Task Source& Task Types &Code Use & Tool Use & Open End & Repository-Level \\ \hline

 Minedojo~\citep{fan2022minedojobuildingopenendedembodied}& -& Internet& Action& \cmark& \cmark& \cmark&\xmark\\
 OSWorld~\citep{OSWorld}& -& Internet& Action& \xmark& \cmark& \cmark&\xmark\\
           ToolBench~\citep{qin2023toolllm}&                 -&                     Tool&            QA&\xmark&          \cmark&          \xmark&                  \xmark\\
           MetaTool~\citep{huang2024metatoolbenchmarklargelanguage}&                 -&                     Tool&            QA&\xmark&          \cmark&          \xmark&                  \xmark\\
 AgentBench~\citep{liu2023agentbenchevaluatingllmsagents}& -& Tool& QA& \cmark& \cmark& \xmark&\xmark\\
 GTSM8K~\citep{cobbe2021trainingverifierssolvemath}& 1& Domain & QA& \xmark& \cmark& \xmark&\xmark\\
 ScienceQA~\citep{lu2022learnexplainmultimodalreasoning}& 3& Domain & QA& \cmark& \xmark& \xmark&\xmark\\
          SciEval~\citep{sun2023scievalmultilevellargelanguage}&                 3&                     Domain &            QA&\xmark&          \xmark&          \xmark&                  \xmark\\
          SciBench~\citep{wang2024scibenchevaluatingcollegelevelscientific}&                 3&                     Domain &            QA&\cmark&          \xmark&          \xmark&                  \xmark\\
          SWE-Bench~\citep{jimenez2024swebench}&                 1&                     GitHub&            Coding&\cmark&          \cmark&          \xmark&                  \cmark ($12$)\\
          ML-Bench~\citep{tang2024mlbenchevaluatinglargelanguage}&                 1&                     GitHub&            Coding&\xmark&          \cmark&          \xmark&                  \cmark ($14$)\\
 SUPER~\citep{bogin2024superevaluatingagentssetting}& -& GitHub& QA& \cmark& \cmark& \xmark&\cmark ($45$)\\ \hline
          \bench (Ours)&                 7&                     Domain and Github&            QA and Coding&\cmark&          \cmark&          \cmark&                  \cmark ($21$)\\ \bottomrule
\end{tabular}
}
\caption{Comparison of benchmarks for evaluating LLMs on domain knowledge and tool utilization. The ``Domain Num.'' column indicates the number of domains evaluated by each benchmark, with ``-'' denoting benchmarks that do not assess domain knowledge. ``Open End'' denotes the presence of an open-ended environment for exploration within the benchmark. ``Repository-Level'' specifies whether the tasks in the benchmark are scoped at the repository level, with the number in the bracket denoting the number of repositories relevant to the benchmark.}
\label{tab:llm-benchmark-comparison}
\end{table*}

\section{Related Work}

\paragraph{LLM-based Agents.}

Large Language Models~(LLMs)~\citep{openaichatgptblog, openai2023gpt4, touvron2023llama,touvron2023llama2} have demonstrated remarkable proficiency across traditional natural language processing~(NLP) tasks. 
LLM-based agents~\citep{autogpt,wu2023autogen,li2023camel,xagent2023, wang2024toolgen, sumers2023cognitive} are LLMs equipped with external tools that can accomplish tasks requiring complex calculations or real-time actions~\citep{NEURIPS2022_82ad13ec, cheng2024seeclick, park2023generative, ye2023proagentroboticprocessautomation, ma2024llm, Cai2023LargeLM, wangexecutable, bogin2024superevaluatingagentssetting, Kumar2023MyCrunchGPTAC, liumind, bran2023chemcrow, huang2024crispr, qi2024metascientisthumanaisynergisticframework, kraus2023enhancing, Koldunov2024, thulke2024climategpt, Vaghefi2023}.
However, existing research typically supports a limited set of tools, which cannot meet the diverse demands of humans. Recently, there has been a focus on tool creation~\citep{cai2023large,qian2023creator,wang2023voyager,qian2023communicative} for agents to dynamically create tools, which are typically file-level code scripts. The functionalities of these created tools remain simple and limited, insufficient to meet real-world open-domain user queries.

\paragraph{Benchmarking LLMs on Domain Knowledge, Tool Use and Open-Domain Tasks.} Different benchmarks evaluate LLMs across diverse domains and capabilities. Domain knowledge benchmarks initially focused on mathematics~\citep{cobbe2021trainingverifierssolvemath, hendrycks2021measuringmathematicalproblemsolving}. Subsequent works\citep{lu2022learnexplainmultimodalreasoning, sun2023scievalmultilevellargelanguage} broadened the scope to encompass three domains: mathematics, physics, and chemistry. \cite{wang2024scibenchevaluatingcollegelevelscientific} further advanced this approach by incorporating code interpreter functionality while continuing to focus on these three domains. These benchmarks are typically derived from established knowledge sources such as textbooks and curated problem repositories, which do not fully capture real-world complexities or cutting-edge questions in rapidly evolving fields.

In parallel, tool use datasets~\citep{qin2023toolllm, huang2024metatoolbenchmarklargelanguage, liu2023agentbenchevaluatingllmsagents} are designed based on the functionalities of pre-defined tools and APIs. 
Recent works have begun to bridge the gap between domain knowledge and practical application by focusing on coding tasks derived from real-world GitHub repositories. However, their scope remains limited to specific domains~\citep{jimenez2024swebench, tang2024mlbenchevaluatinglargelanguage, bogin2024superevaluatingagentssetting}.

Lastly, while open-domain exploration is crucial for real-world tasks, existing research has primarily studied it in action-oriented environments~\citep{fan2022minedojobuildingopenendedembodied, wang2023voyager, OSWorld}.

In conclusion, existing benchmarks remain limited in their scope, domains, tool use, or coding tasks in isolation. 
%
%
They cannot be utilized to evaluate the open-domain task-solving capability of LLMs.
Table~\ref{tab:llm-benchmark-comparison} lists the main differences between our benchmark and previous works.

%% file: section/dataset.tex
\section{\bench}
\label{sec:dataset}

\begin{figure*}[!ht]
\centering
\includegraphics[width=\linewidth]{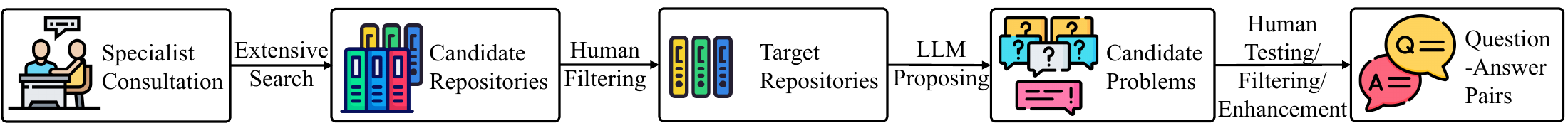}
\caption{The construction pipeline of \bench.}
\label{fig:con}
\end{figure*}  

\subsection{Dataset Construction}
\label{sec: datacon}

We present \bench, a high-quality benchmark spanning $7$ professional domains that bridge open-domain knowledge with executable implementation resources. As is shown in Figure~\ref{fig:con}, the dataset construction process involves the following stages:

\textbf{Domain-Specific Problem Curation.} We collaborated with domain experts to identify frontier challenges within their respective fields that are amenable to computational solutions. This process yielded 7-10 candidate problems per domain through iterative refinement with expert feedback.

\textbf{Repository Selection and Filtering.} We conducted systematic searches on GitHub to identify repositories implementing solutions for the curated problems. From an initial pool of 10 repositories per domain, we applied rigorous filtering criteria: (1) Removal of repositories with duplicate functionality or implementation approaches; (2) Verification of active maintenance status within the past 6 months; (3) Assessment of documentation quality. This yielded 21 repositories across seven domains.

\textbf{Query Generation and Validation.} For each repository, we employed GPT-4 to generate 30 candidate queries of varying complexity levels. Through manual validation, we retained 5-10 executable queries per repository, ensuring: (1) Solvability using the target repository; (2) Coverage of different functionality aspects; (3) Absence of ambiguous phrasing. This process resulted in 113 validated base queries.

\textbf{Query Augmentation.} To investigate the capability of LLMs to search for the proper repositories, we designed three prompt conditions for each base query: (1) \textit{Explicit Hint}: Direct repository specification with GitHub URL; (2) \textit{Implicit Hint}: Domain/keyword-based hints about the repository without specific identification; (3) \textit{No Hint}: Base query without any supplemental information about the repository.

This query augmentation approach yields 339 instruction-answer pairs with expert-validated ground truth solutions. Table~\ref{tab:query_stat} summarizes the dataset statistics across domains. Appendix~\ref{sec:appendix_prompt} contains more detailed statistics of \bench.

\begin{table}[!b]
    \centering
    \small
    \begin{tabular}{lrr}
        \toprule
        Domain & \multicolumn{1}{c}{Num. of Repo.} & \multicolumn{1}{c}{Num. of Query} \\
        \hline
        Finance& $2$ & $45$ \\
        Chemistry& $4$ & $66$ \\
        Bioinformatics& $2$ & $30$ \\
        Computer Vision& $6$ & $90$ \\
        Network Analysis& $2$ & $30$ \\
        Security Analysis& $2$ & $30$ \\
        Visualization& $3$ & $48$ \\
        \hline
        Total & $21$ & $339$ \\
        \bottomrule
    \end{tabular}
    \caption{Statistics of our constructed \bench.}
    \label{tab:query_stat}
\end{table}

\subsection{Data Categorization}
\label{sec:categorization}

We categorized the collected repositories based on the difficulty of the Setup and Apply phases.

For the \textbf{Setup difficulty}, we divided the collected repositories into $3$ classes:
(1)\textit{Setup-Easy}: The README provides a detailed and correct setup tutorial, with which the environment can be set up fluently.
(2)\textit{Setup-Medium}: The README misses some details or contains slight flaws, which require the agent to solve based on error reports.
(3)\textit{Setup-Hard}: The README provides an incorrect tutorial because of human error or insufficient maintenance, which needs agents to find relevant Issues/PRs to solve.

For the \textbf{Apply difficulty}, we also divided the repositories into $3$ classes:
(1)\textit{Apply-Easy}: Simply requires running some commands given by the README.
(2)\textit{Apply-Medium}: Requires writing configuration files or downloading extra resources, like data and trained models.
(3)\textit{Apply-Hard}: Requires modifying the source code of the repositories or referring to relevant Issues/PRs for help.

\subsection{Operation Environment}

To ensure experiment reproducibility and minimize dependency on local environments, we designed a comprehensive interface that allows LLMs to interact seamlessly with a Docker container. In our work, all interactions with the LLMs are executed within a controlled Docker environment.

\subsection{Evaluation Metrics}

We designed 2 evaluation metrics for tasks in \bench: Completeness and Pass Rate.

\paragraph{Pass Rate}
The Pass Rate is defined as the proportion of queries that successfully meet the predefined criteria relative to the total number of queries. This evaluation is conducted exclusively based on the comparison of the final answer with the expert-generated "golden answers" by GPT-4. A query is deemed to pass if there is a concordance between these two answers.

\paragraph{Completeness} To precisely evaluate the performance of \model and its baselines, we further designed a metric to evaluate the whole execution process with a GPT-4-based evaluation agent, scoring from $0$ to $10$. The evaluation covers three phases: Search, Setup, and Apply. GPT-4 assigns scores of $[0, 3]$ for Search, Setup, and Apply, and $[0, 1]$ for the final answer. These scores are subsequently aggregated and normalized to a 10-point scale to derive the overall completeness score.

We sampled $120$ queries and results for both GPT-4 evaluation and human annotation, achieving an $87.5$\% absolute match, indicating the high reliability of our metrics. The details of this check are depicted in Appendix~\ref{ac}.

%% file: section/method.tex
\section{\model}

Our preliminary experimental results show that both vanilla LLMs and general-purpose LLM agents perform poorly on \bench. Even when we provide the LLM with relevant GitHub interfaces and a well-constructed environment to create a custom LLM agent, the model's performance remains inferior.
We attribute it to several challenges in employing GitHub repositories to fulfill a task:
(1) \textbf{Lack of Quality Assurance}: GitHub repositories often lack standardization and may contain flaws or bugs, and their documentation may also be incomplete, misleading, or contain errors.
(2) \textbf{Alignment Gap between Tools and Queries}: Tools on GitHub are not specifically designed for given queries, so they need adjustments to suit the users' needs.
%
(3) \textbf{Workflow Complexity}: The whole workflow involves dozens of different tasks. The significant differences between these tasks can easily distract the LLMs from completing the whole process effectively.

To address this, we propose \model, a LLM agent for open-domain task-solving with $2$ novel features: (1) Hierarchical Agent System, where models delegate subtasks to reduce workflow burden~(Section~\ref{sec:has}), (2) Bi-Level Experience Learning~(Section~\ref{sec:experience-learning}), which learns from Issues/PRs for in-task knowledge and accumulates cross-task experience. The first layer of the Hierarchical Agent System consists of three main components, and we introduce them in Section~\ref{mps}.

\subsection{Hierarchical Agent System}
\label{sec:has}

\begin{figure}[!ht]
\small
    \centering
    \includegraphics[width=.8\linewidth]{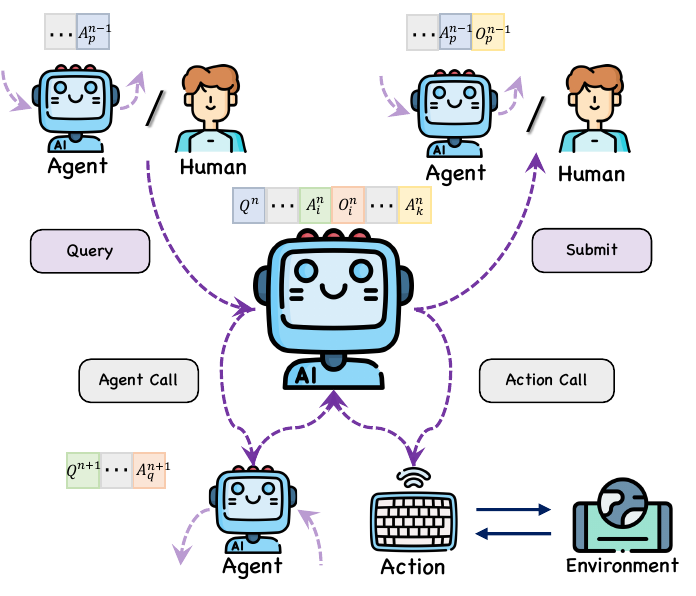}
    \caption{Illustration of the Hierarchical Agent System, where blocks mean memory list and the same background color denotes the same information.}
    \label{fig:aagnecy}
\end{figure}

A Hierarchical Agent System breaks down complex tasks into smaller sub-tasks, where each agent receives a query from a higher-level entity and responds by either taking actions to interact with the environment or delegating tasks to sub-agents. In other words, the actions of agents in a hierarchical agent system fall into two categories: (1) Action Calls: Direct interactions with the environment, such as executing commands, checking files, or submitting results. (2) Agent Calls: Assigning sub-tasks to specialized agents, such as setting up environments or modifying files.

Each agent processes its query based on previous actions and observations. If it delegates to a sub-agent, the sub-agent continues the process with a refined query. If the agent acts, it either completes the task (reporting the result back) or interacts with the environment and records feedback.

This recursive process continues until all tasks are broken down into simple, executable actions. By structuring tasks hierarchically, the system efficiently manages complexity, allowing each agent to focus on its specific role while collectively achieving intelligent behavior.

\begin{figure*}[!ht]
\centering
\includegraphics[width=\linewidth]{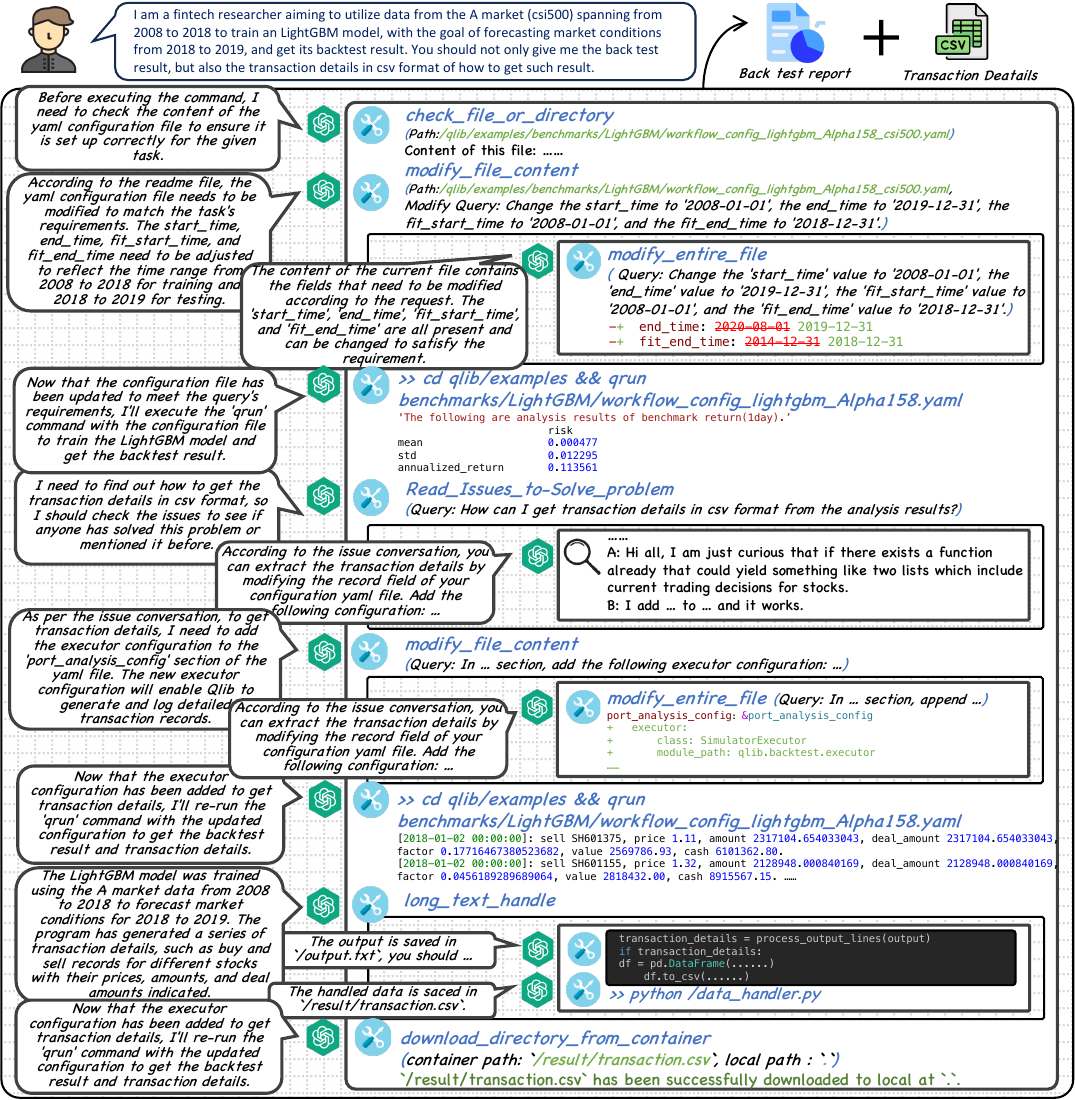}
\caption{An example of the tool application of Qlib, where the agent called $4$ sub-agents to fulfill the given task.}
\label{fig:case_apply_qlib}
\end{figure*}

Formally, we use $Agent_k^n$ to denote the $k$-th agent at level $n$ of the hierarchy, $A_i^n$ denotes the $i$-th action, which can be tool-using or designating inferior agents, by $Agent^n$. When $Agent_k^n$ receives query $Q^n$ from its superior agent or human, the problem solving process can be formulated as $A_i^n = Agent_k^n(Q^n, A_j^n, O_j^n,\ldots,A_1^n,O_1^n)$, where $O_j^n$ and $A_j^n$ are respectively the observations and preceding actions that lead up to $A_i^n$. If $A_i^n$ is calling a sub-agent, the query $Q^{n+1}$ for $Agent_k^{n+1}$ is derived from $A_i^n$. When $Agent_k^{n+1}$ finishes its task, it will report the result $A_q^{n+1}$ to $Agent_k^n$. Figure~\ref{fig:aagnecy} demonstrated this hierarchical recursive process.

\subsection{Bi-Level Experience Learning}
\label{sec:experience-learning}

We developed and implemented an experience learning feature for \model, encompassing both in-task and cross-task learning paradigms.

Due to the non-standardization of GitHub repositories, some lack perfect READMEs and necessary setup information. Additionally, flaws in the source code pose challenges. In such cases, learning from human experiences becomes an efficient approach.

Building upon the hierarchical framework, we introduce a specialized agent, the Issue/PR Agent to handle the experience learning process. This agent is called when a higher-level agent encounters a problem that might benefit from past experiences or community solutions. It is responsible for searching, evaluating, and returning relevant information from GitHub Issues and Pull Requests.

Apart from in-task knowledge learned from community experiences and solutions, \model can also learn from its own past experiences to improve decision-making over time. It updates its knowledge by summarizing past actions and outcomes, refining its approach for future tasks.

\begin{table*}[t]
\centering
\resizebox{\textwidth}{!}{
\begin{tabular}{c c c c c c c c c}
\toprule
\textbf{Methods} & Finance & Chemistry & Bioinformatics & Computer Vision & Network Analysis & Security Analysis & Visualization & \textbf{Avg.} \\ 
\midrule
\multicolumn{9}{c}{\textbf{\textit{GPT-3.5-Turbo Based}}} \\ 
\midrule
Vanilla & 0.0 & \textbf{36.4} & 0.0 & 0.0 & 0.0 & 0.0 & \textbf{31.3} & 11.5 \\ 
ReAct & 2.2 & 3.0 & 3.3 & 6.7 & 0.0 & 0.0 & 0.0 & 2.4 \\ 
ReAct + Sum. & 0.0 & 0.0 & 0.0 & 0.0 & 0.0 & 0.0 & 0.0 & 0.0 \\ 
\textbf{\model(Ours)} & \textbf{8.9} & 24.2 & \textbf{23.3} & \textbf{8.9} & \textbf{10.0} & \textbf{33.3} & 20.1 & \textbf{17.1} \\ 
\midrule
\multicolumn{9}{c}{\textbf{\textit{GPT-4 Based}}} \\ 
\midrule
Vanilla & 0.0 & \textbf{68.2} & 0.0 & 0.0 & 0.0 & 0.0 & 43.8 & 19.5 \\ 
XAgent & 0.0 & 40.9 & 0.0 & 0.0 & \textbf{40.0} & 0.0 & \textbf{81.3} & 23.0 \\ 
ReAct & 51.1 & 19.7 & 17.8 & 22.2 & 10.4 & 30.0 & 23.3 & 24.6 \\ 
ReAct + Sum. & 31.1 & 19.7 & 26.7 & 22.9 & 14.8 & 33.3 & 26.7 & 24.4 \\ 
\textbf{\model(Ours)} & \textbf{68.9} & 34.9 & \textbf{86.7} & \textbf{45.6} & 16.7 & \textbf{43.3} & 35.4 & \textbf{47.3} \\ 
\bottomrule
\end{tabular}
}
\caption{Pass Rates ($\%$) of different methods across various domains in the \bench dataset. Results are shown for both GPT-3.5-Turbo and GPT-4-based implementations. ``Avg.'' represents the average pass rate across all domains.}
\label{tab:main}
\end{table*}

After completing a task, the agent stores its execution environment in a Docker image, allowing for easy reuse when similar queries arise. To enhance retrieval, it abstracts the repository’s functionality and summarizes key experiences, ensuring efficient adaptation to new challenges.

\subsection{Main Phases}
\label{mps}

\paragraph{Repository Search.}

In the search phase, the agent identifies suitable repositories to fulfill the user's query. It first checks previously stored repositories to determine if any match the query. If a suitable repository is found, its environment is loaded directly, bypassing the setup phase. If no stored repository is available, the agent searches GitHub for relevant options. If the user specifies a repository, the agent retrieves it by name; otherwise, it searches by topic to find relevant repositories. Once candidate repositories are gathered, the agent evaluates their suitability by analyzing their README files and determining whether they can effectively address the user’s request. 

\paragraph{Environment Setup.}

Once a suitable repository is identified, the agent sets up its execution environment. This begins with cloning the repository and installing dependencies as outlined in its README file. However, since many repositories lack standardized documentation or contain flaws, the agent may need to search GitHub Issues and Pull Requests to resolve problems. If necessary, it modifies the repository’s source files to fix bugs. To ensure security and isolation, all operations are conducted within a Docker environment.

\paragraph{Tool Application.}

After configuring the environment, the agent proceeds to apply the repository to solve the user query. If the repository is well-structured and provides a clear entry point, such as a command-line interface, the agent can use it directly. For non-standardized repositories with limited documentation, the agent relies on human experiences extracted from Issues and PRs. If the execution produces extensive output, the agent writes a Python script to extract key information efficiently. Figure~\ref{fig:case_apply_qlib} shows an example of \model fulfilling a task by interacting with the environment and assigning tasks to lower agents.

An elaborated introduction to these phases is in Appendix~\ref{ela}. Note that although this process follows a hierarchical structure, the agent dynamically decides which phase, subtask, or action to execute based on the specific query. This flexibility ensures adaptability to diverse and complex tasks.

%% file: section/exp.tex
\section{Experiment}

\subsection{Experiment Settings}

\paragraph{Baselines.}
To validate the effectiveness of our \model, we design the following baselines:
(1) LLM: Vanilla LLMs without external tools
(2) ReAct~\citep{yao2022react}: ReAct is a widely-used LLM-based agent task-solving technique~\citep{autogpt,wu2023autogen}.
In our settings, ReAct is equipped with the same actions as our \model to extend tools from GitHub for fair comparison.
(3) ReAct+Summary: Due to the complexity of the tool extension, the whole process tends to involve lengthy context, surpassing the context window of LLMs.
Hence, we design this ReAct variant which will summarize the context when the length of the context reaches the threshold.
(4) XAgent~\citep{xagent2023}: XAgent is a powerful general-purpose LLM-based agent, which is equipped with numerous external tools and can reason, plan, code, and reflect.

\paragraph{Implementation Details.}
We implement \model and baseline methods except XAgent based on the \texttt{gpt-4-0125-preview} and \texttt{gpt-3.5-turbo-16k} respectively with a $0.7$ temperature, under a $0$-shot setting. There is no GPT-3.5-based XAgent because its reasoning and planning ability can't support XAgent's complex workflows.
%
%
Specially, the following main experiments reflect the success rate on the first encounter with the problem, incorporating PR/Issue but without using experience summary~(cross-task experience learning) to enhance the strategy.

\subsection{Overall Results}


Table~\ref{tab:main} reports the Pass Rates of each method. We get several observations:
(1) While Vanilla LLMs and XAgent demonstrate good performance in familiar domains like Chemistry and Visualization, it is hard for them to answer questions in unacquainted domains like Bioinformatics, Finance, etc.
(2) ReAct achieves a lower Pass Rate than the agency structure in both settings, which demonstrates that simply adapting the ReAct framework cannot achieve good results.
(3) ReAct+Summary achieves lower performance than ReAct because the summarization will lose critical information. 
Thus, it is infeasible to avoid the over-length problem by simply summarizing the long context.
(4) All GPT-4-based methods outperform their GPT-3.5 counterparts significantly, showing that \bench is a challenging dataset that requires powerful LLMs to achieve.
(5) \model significantly outperforms all baselines in both settings, demonstrating the effectiveness of the hierarchical agent system.

\subsection{Abalation Study}

\begin{table}[!t]
    \centering
    \small
    \begin{tabular}{lr}
        \toprule
        Method & Pass Rate \\
        \midrule
        \model w/ PRs\&Issues & $47.3$ \\
        \model w/o PRs\&Issues & $40.3$ \\
        \bottomrule
    \end{tabular}
    \caption{Results of ablating in-task experience learning.}
    \label{tab:exp_human_exp}
\end{table}

To evaluate the effectiveness of the bi-level experience learning mechanism, we conduct an ablation study.
For \textbf{in-task experience learning}, we remove the PRs/Issues actions to re-run the main experiments.
Experimental results are shown in Table~\ref{tab:exp_human_exp}, and we observe that without PRs/Issues, the pass rate decreases to $40.3$\%.
It not only verifies the non-standardization problem of GitHub repositories but also proves that learning from PRs/Issues can overcome this challenge.

\begin{table}[!t]
    \centering
    \small
    \begin{tabular}{lrr}
        \toprule
        Method & w/o SelfExp & w/ SelfExp \\
        \midrule
        GPT-3.5 & $17.6$ & $58.8$ \\
        GPT-4 & $47.0$ & $82.3$ \\
        \bottomrule
    \end{tabular}
    \caption{Experimental results of employing cross-task experience learning.}
    \label{tab:exp_self_exp}
\end{table}

For \textbf{cross-task experience learning}, we select $2$ repositories: \texttt{Qlib} and \texttt{AiZynthFinder}, which both belong to the Hard Apply category.
We run their $51$ queries and utilize the GPT-4-based \model to store the repositories with summarized practice experience.
We then re-run these queries but \model would retrieve the stored repositories and utilize the summarized experience to accomplish the queries.
As shown in Table~\ref{tab:exp_self_exp}, leveraging the experience summarized by GPT-4-based \model, GPT-3.5-based \model can even achieve a higher pass rate than GPT-4-based \model without summarized experience.
Simultaneously, GPT-4-based \model can achieve a higher Pass Rate even though it leverages the summarized experience by itself.

The above studies prove the effectiveness of bi-level experience learning.

\subsection{Impact of Different Phases}

We delve deeper into each phase from a thorough understanding of \bench and \model.

\paragraph{Search.} 
As introduced in Section~\ref{sec: datacon}, we design three types of prompts to denote the target repositories.
We calculate the Seach Success Rate of each type of prompt.
If \model can get the correct repositories of a query, it is denoted as search success.
Then, we calculate the proportion of the search success queries over all queries of each type of prompt.
The experimental results are listed in Table~\ref{tab:exp_search}.
We can observe that the Explicit Repo Prompt achieves the highest Search  Success Rate (nearly 100\%) as the prompt has specified repositories.
Implicit Repo Prompt achieves $66.0$\% Search Success Rate, showing that \model can infer the relevant GitHub Topics based on the domains or careers.
Finally, if no repository prompt is provided, the search success rate decreases significantly.
It demonstrates that \model falls short of inferring GitHub Topics based on the query only.
It needs further research in the future to improve the performance in this situation.

\begin{table}[!ht]
  \centering
  \small
      \begin{tabular}{lr}
          \toprule
          \multirow{1}*{Prompt} & \multicolumn{1}{c}{Search Success Rate}  \\
          \midrule
          Explicit Repo Prompt & $96.0$  \\
          Implicit Repo Prompt & $66.0$  \\
          No Repo Prompt & $32.0$ \\
          \bottomrule
      \end{tabular}
  \caption{Analysis for the search difficulty.}
  \label{tab:exp_search}
\end{table}

\paragraph{Setup \& Apply.}

Table~\ref{tab:exp_setup_apply} shows the Pass Rates for the repositories categorized based on Setup and Apply difficulties, as described in Section~\ref{sec:categorization}.

For Setup difficulty, both Medium and Hard repositories achieve similar Pass Rates. It shows that \model's human experience learning capability helps overcome imperfect READMEs.

For Apply difficulty, the Pass Rate for Hard decreases by over $12\%$ compared to Easy and Medium. This demonstrates that while \model can effectively handle repositories with easy and medium Apply difficulty, it requires further study to conquer those with hard Apply difficulty.

\subsection{Error Analysis}

Although our method can autonomously extend tools from GitHub, we still observe some failures.
%

\paragraph{Repository Select Failure.}

A key challenge for \model is selecting the appropriate repository from GitHub to address user queries. 
We noticed instances where \model selected repositories that were not capable of resolving the given queries. 
This issue was particularly prevalent when the user query did not specify a particular repository.
Then the agent's decision-making process relies heavily on the README files of repositories. 
However, these files sometimes lack clear and explicit descriptions of the repository's functionality, or even overclaim it, and lead to misjudgments by the agent. 
For example, in the Finance scenario, \model erroneously selected the \texttt{vnpy} repository, which is suited for quantitative trading but not for research applications like exploring specific models. 

\paragraph{Environment Configuration Failure.}

In cases like \texttt{Bringing-Old-Photos-Back-to-Life}, we observed failures in the environment setup. 
While an official Dockerfile was present, it was outdated and non-functional without enough maintenance. 
The correct Dockerfile was located within a Pull Request, which the agent should ideally access to find the accurate setup instructions.
However, \model sometimes opted to modify the existing, incorrect Dockerfile rather than seeking the correct version in PRs. 
Due to unresolved bugs in the Dockerfile, the agent was unable to correctly set up the environment, leading to failure.

\paragraph{Execution Configuration Failure.}

The repository \texttt{Qlib} presented unique challenges, as it requires writing a specific configuration file for execution. 
This file encompasses a range of parameters, including dataset settings, model hyperparameters, and backtesting parameters. 
Incorrect settings in any of these parameters can lead to results that do not meet the user queries' requirements.
In practice, we observed that \model may incorrectly set the time range for data or specify erroneous file paths, resulting in execution failure.

\begin{table}[!t]
  \centering
  \small
  \resizebox{1.0\linewidth}{!}{
      \begin{tabular}{lrrrr}
          \toprule
          \multirow{1}*{Setup/Apply Difficulty} & \multicolumn{1}{c}{Easy} & \multicolumn{1}{c}{Medium} & \multicolumn{1}{c}{Hard} & \multicolumn{1}{c}{Total} \\
          \midrule
          Easy & $72.3$ & $69.0$ & $56.2$ & $64.4$ \\
          Medium & $60.7$ & $70.0$ & $41.5$ & $57.7$ \\
          Hard & $50.0$ & $67.0$ & $51.5$ & $57.4$ \\
          \midrule
          Total & $64.1$ & $68.7$ & $51.4$ & $60.7$ \\
          \bottomrule
      \end{tabular}
  }
  \caption{Analysis for the setup \& apply difficulty.}
  \label{tab:exp_setup_apply}
\end{table}

\paragraph{General Failure Cases of Agents.}

In some domains like visualization, vanilla models outperformed agentic LLMs, including XAgent and \model. Case studies revealed that forcing tool use for questions that LLMs can already answer correctly may lead to inferior performance due to incorrect interactions. Recent studies have widely observed similar failure cases in LLM Agent Systems~\citep{yu2024tooling, cemri2025multi}. In domains unfamiliar to LLMs (e.g., quantitative finance), \model significantly outperforms vanilla LLMs and general-purpose LLM Agents. It indicates our methods possess great potential in open-domain problem-solving.


We also conduct a Case Study in \Cref{sec:case_study} to further show the detailed process of \model.

%% file: section/conclusion.tex
\section{Conclusion}
In this paper, we introduced \bench, a comprehensive benchmark designed to evaluate the open-domain task-solving capabilities of LLMs. Our experiments highlighted the limitations of existing LLM-based agents and demonstrated the effectiveness of our proposed OpenAgent system. OpenAgent's hierarchical framework and bi-level experience learning mechanism significantly enhance its capabilities, allowing them to tackle complex tasks across diverse domains.
Our work paves the way for more robust and flexible LLM-based agents, capable of evolving alongside rapidly changing technological landscapes.

%% file: section/limitation.tex
\section{Limitation}
Our study has explored the tool extension capability of LLM-based agents, yet there exist certain limitations and risks.
Firstly, our method relies on the utilization of Pull Requests~(PRs) and Issues from GitHub as primary sources of human experience. 
However, it is important to recognize that similar functionalities may not be universally available across other repository hosting platforms. 
Consequently, the generalizability of our findings is confined to the GitHub ecosystem, posing a limitation to the applicability of our approach beyond this specific context.
Secondly, the dynamic nature of GitHub repositories, characterized by frequent updates, bug fixes, and the evolution of repository functionalities, introduces a layer of volatility. 
This fluidity can significantly impact the reproducibility of our experimental results over time, as the state of the repositories at the time of study may not reflect their future states.
Thirdly, our method necessitates the use of Docker for the execution of repository content. 
This dependency on Docker implies substantial consumption of server resources, including CPU, memory, and storage. 
Such resource-intensive requirements may pose practical limitations on the scalability and feasibility of deploying \model, particularly in environments with constrained computational resources.

\section*{Acknowledgements}
This work was supported in part by the National Natural Science Foundation of China (Grant No. 62376273), the Postdoctoral Fellowship Program of CPSF (Grant No. GZB20230343 and Grant No. GZC20240831) and the China Postdoctoral Science Foundation (Grant No. 2023M741945).

We thank all collaborators who made their contributions to this project, including
Bohan Lyu (\email{lvbh22@mails.tsinghua.edu.cn}),
Xin Cong (\email{xin.cong@outlook.com}),
Heyang Yu (\email{hy-yu22@mails.tsinghua.edu.cn}),
Pan Yang (\email{im.panyang@gmail.com}),
Cheng Qian (\email{chengq9@illinois.edu}),
Zihe Wang (\email{wangzihe18@mails.ucas.ac.cn}),
Yujia Qin (\email{yujiaqin16@gmail.com}),
Yining Ye (\email{yeyn23@mails.tsinghua.edu.cn}),
Chen Qian (\email{qianc@sjtu.edu.cn}),
Zhong Zhang (\email{zhongzhang@tsinghua.edu.cn}),
Yukun Yan (\email{yanyk.thu@gmail.com}),
Yankai Lin (\email{yankailin@ruc.edu.cn}),
Zhiyuan Liu (\email{liuzy@tsinghua.edu.cn}),
and Maosong Sun (\email{sms@tsinghua.edu.cn}).
For further information or collaboration, feel free to contact them.

%% file: section/appendix.tex

\section*{Appendix}

\section{Prompt Details}\label{sec:appendix_prompt}

\input{section/prompts.tex}


\section{GPT-4 Evaluation Alignment Experiment}
\label{ac}

\begin{figure}[!t]
    \centering
    \includegraphics[width=\linewidth]{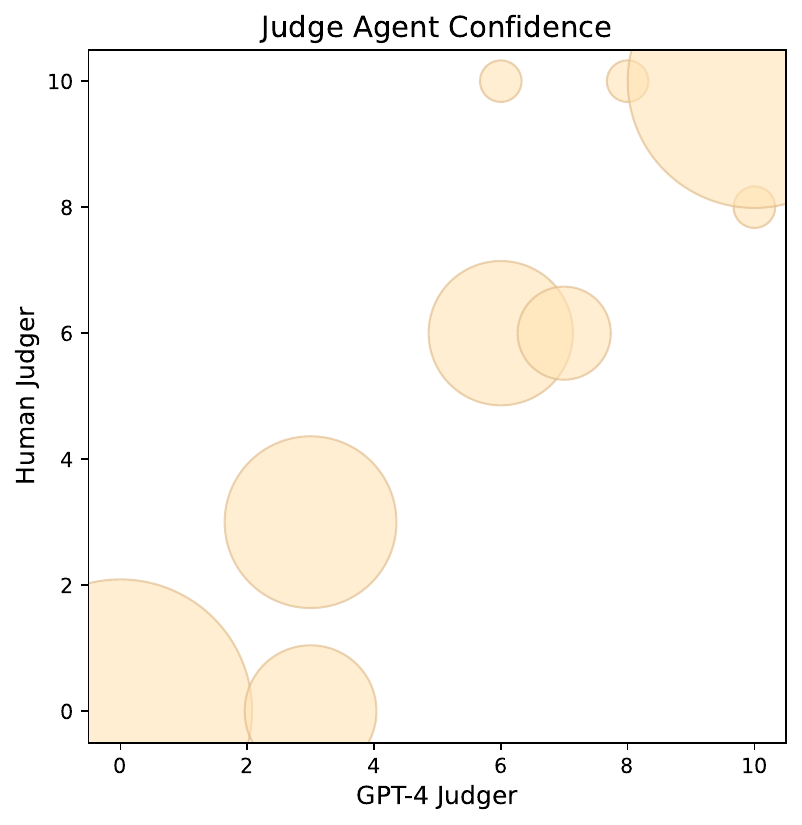}
    \caption{GPT-4 Evaluation Alignment Experiement}
    \label{fig:ac}
\end{figure}

We randomly selected 120 questions from the \bench and conducted both manual scoring and machine scoring using our designed GPT-based Agent, then visualized the results in the above figure. The circular points represent the coordinates corresponding to human scores and GPT-4 scores, with the size of the circles indicating the number of questions with that particular score combination. As shown, there is a high consistency between GPT-4 scoring and human assessment.


\section{Dataset Details}
\label{ap:b}

Table~\ref{table:github_repos} provides the overall statistics regarding the repositories. 
Table~\ref{table:github_repos_by_field} provides the field of each Github repository.
Table~\ref{tab:category} provides the difficulty of each Github repository.

\begin{table*}[!t]
\centering
\begin{tabular}{p{3.4cm}p{3.4cm}p{7.2cm}}
\toprule
\textbf{Author} & \textbf{Name} & \textbf{Address} \\ \hline
danielgatis & rembg & \url{https://github.com/danielgatis/rembg} \\
ocrmypdf & OCRmyPDF & \url{https://github.com/ocrmypdf/OCRmyPDF} \\
cdfmlr & pyflowchart & \url{https://github.com/cdfmlr/pyflowchart} \\
HarisIqbal88 & PlotNeuralNet & \url{https://github.com/HarisIqbal88/PlotNeuralNet} \\
lukas-blecher & LaTeX-OCR & \url{https://github.com/lukas-blecher/LaTeX-OCR} \\
s0md3v & Photon & \url{https://github.com/s0md3v/Photon} \\
s0md3v & Bolt & \url{https://github.com/s0md3v/Bolt} \\
s0md3v & Smap & \url{https://github.com/s0md3v/Smap} \\
MultiQC & MultiQC & \url{https://github.com/MultiQC/MultiQC} \\
xinyu1205 & recognize-anything & \url{https://github.com/xinyu1205/recognize-anything} \\
bukosabino & ta & \url{https://github.com/bukosabino/ta} \\
molshape & ChemFormula & \url{https://github.com/molshape/ChemFormula} \\
tencent-quantum-lab & TenCirChem & \url{https://github.com/tencent-quantum-lab/TenCirChem} \\
harirakul & chemlib & \url{https://github.com/harirakul/chemlib} \\
ultralytics & yolov5 & \url{https://github.com/ultralytics/yolov5} \\
mermaid-js & mermaid-cli & \url{https://github.com/mermaid-js/mermaid-cli} \\
microsoft & qlib & \url{https://github.com/microsoft/qlib} \\
fritzsedlazeck & Sniffles & \url{https://github.com/fritzsedlazeck/Sniffles} \\
MolecularAI & aizynthfinder & \url{https://github.com/MolecularAI/aizynthfinder} \\
microsoft & Bringing-Old-Photos-Back-to-Life & \url{https://github.com/microsoft/Bringing-Old-Photos-Back-to-Life} \\
PyCQA & bandit & \url{https://github.com/PyCQA/bandit} \\ \bottomrule
\end{tabular}
\caption{GitHub Repositories}
\label{table:github_repos}
\end{table*}



\begin{table}[!t]
\centering
\resizebox{\linewidth}{!}{
\begin{tabular}{ll}
\toprule
\textbf{Domain} & \textbf{Repository} \\ \midrule
\multirow{2}{*}{Finance} & microsoft/qlib \\
& bukosabino/ta \\
\midrule
\multirow{4}{*}{Chemistry} & molshape/ChemFormula \\
& tencent-quantum-lab/TenCirChem \\
& harirakul/chemlib \\
& MolecularAI/aizynthfinder \\
\midrule
\multirow{2}{*}{Bioinformatics} & MultiQC/MultiQC \\
& fritzsedlazeck/Sniffles \\
\midrule
\multirow{6}{*}{CV} & danielgatis/rembg \\
& lukas-blecher/LaTeX-OCR \\
& ultralytics/yolov5 \\
& microsoft/Bringing-Old-Photos-Back-to-Life \\
& mermaid-js/mermaid-cli \\
& xinyu1205/recognize-anything \\
\midrule
\multirow{2}{*}{Network Analysis} & s0md3v/Photon \\
& s0md3v/Smap \\
\midrule
\multirow{2}{*}{Security Analysis} & PyCQA/bandit \\
& s0md3v/Bolt \\
\midrule
\multirow{3}{*}{Chart Paint} & cdfmlr/pyflowchart \\
& ocrmypdf/OCRmyPDF \\
& HarisIqbal88/PlotNeuralNet \\
\bottomrule
\end{tabular}
}
\caption{GitHub repositories categorized by $7$ fields.}
\label{table:github_repos_by_field}
\end{table}




\section{Details about Main Phases of \model}
\label{ela}

\paragraph{Repository Search}

During the Search phase, the agent finds suitable repositories that can be used to accomplish user queries.
The repositories come from two resources: repositories stored in the past and repositories hosted in GitHub.
Hence, this phase contains three subtasks:
(1) Stored Repository Retrieval: 
The agent retrieves from existing stored repositories by judging their suitability with the user query. 
%
%
If a repository is deemed suitable, its environment is loaded, bypassing the subsequent Setup phase, and directly enters the Apply phase.
(2) GitHub Repository Search:
If the stored repositories cannot be used to accomplish user queries, the agent will resort to GitHub to search for suitable ones.
There are two ways to search for repositories.
If the user queries specify the particular repositories, the agent will take action to call GitHub \textit{search by name} API directly.
If not, \model should search for the proper repositories according to the repository function.
As GitHub lacks the semantic search API, we resort to the topic search API.
The agent would extract a list of potential GitHub topics from the query and subsequently call GitHub \textit{search by topic} API to search repositories.
(3) Repository Function Judgment:
Upon obtaining repository candidates, the agent judges each repository's suitability in resolving the user query. 
The agent will read the README of each repository to understand its function and then deliver a judgment on the repository's suitability.
%

\paragraph{Environment Setup}

Upon identifying the suitable repositories, the agent would initiate the \textit{Setup} phase aimed at configuring their execution environment.
The agent commences by cloning repositories from GitHub and executing commands (including the installation of dependencies and download of requisite data) according to the README. 
Due to the non-standardization problem, there may exist flaws or bugs in the repositories so the agent will initiate a \textit{Pull Requests Exploration} or \textit{Issues Exploration} subtask to leverage human practice experience to resolve the problems. 
If necessary, the agent will initiate a \textit{File Modification} subtask to modify the source files to fix the bugs.
%

\paragraph{Tool Application}

Given the configured environment, the agent proceeds to apply the repository to address the user query. 
This application process varies based on the complexity and design of individual repositories.
Well-developed repositories provide clear entry for allowing straightforward applications (e.g., Command-Line Interface). 
%
%
Nevertheless, for those non-standardized repositories that do not provide clear entry, especially lacking detailed documentation, the agent needs to resort to human experience again (see in Section~\ref{sec:experience-learning}).
If extensive output (e.g., lengthy execution logs) ensues, the agent needs to go to the \textit{Long Context Process} subtask which writes a Python program (e.g., regular expressions) to extract critical information from the lengthy file.
Thus, the \textit{File Modification} subtask is also involved.

Note that although we design this hierarchical strategy, which phase, subtask, or action to be achieved is decided by \model itself dynamically.
We do not limit the agent's behavior strictly.


\section{Case Study}
\label{sec:case_study}

To detail how \model works during the whole tool extension process, we conduct the case study to demonstrate the behavior of \model.

\begin{figure*}[!ht]
  \centering
  \includegraphics[width=1.0\linewidth]{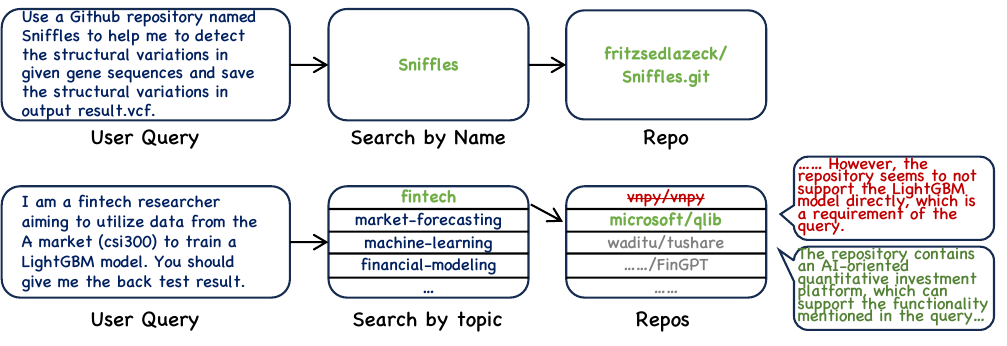}
  \caption{Comparison of two different repository search methods.}
  \label{fig:case_search_comparison}
\end{figure*}

\begin{figure*}[!ht]
\centering
\includegraphics[width=0.9\linewidth]{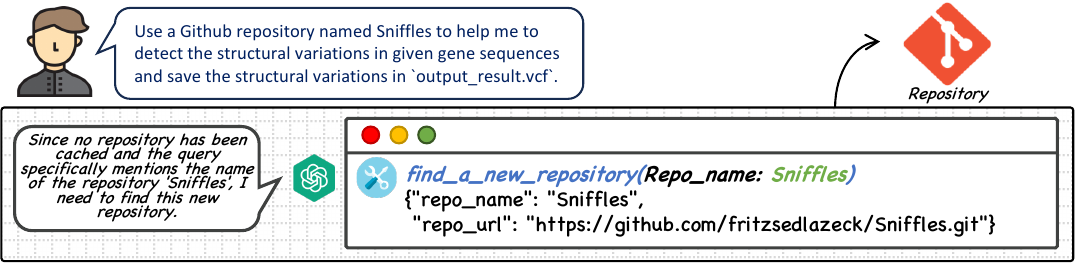}
\caption{An example about the repository search of \texttt{Sniffles}.}
\label{fig:case_search_sniffles}
\end{figure*}

\begin{figure*}[!t]
\centering
\includegraphics[width=0.9\linewidth]{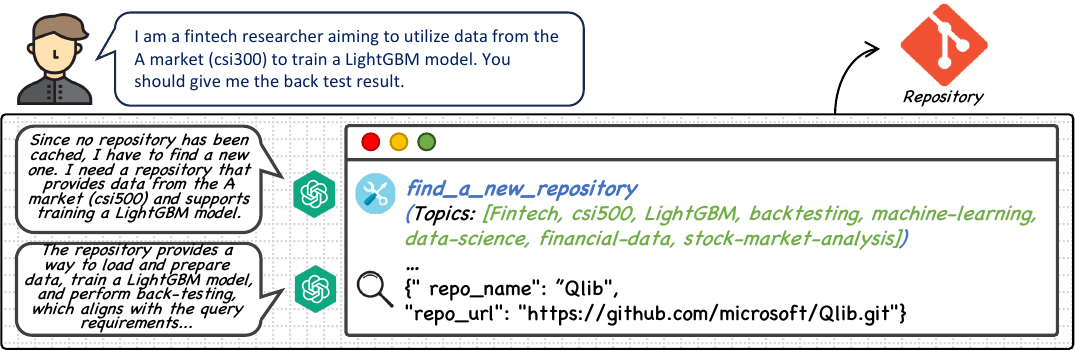}
\caption{An example about the repository search of \texttt{Qlib}.}
\label{fig:case_search_qlib}
\end{figure*}

\textbf{Adaptive Repository Search Strategies.}
\model demonstrates a remarkable ability to autonomously select and implement varied search strategies for repository retrieval (see in Figure~\ref{fig:case_search_comparison}). 
This adaptability is evident from its high search success rate across different repositories. 
\model tailors its search approach based on the specificity of the user query. 
For instance, in the case of \texttt{Sniffles}, where the repository name is provided (Figure~\ref{fig:case_search_sniffles}), \model directly searches for the repository using the given name. 
In contrast, for queries of Qlib, where no specific repository is mentioned (Figure~\ref{fig:case_search_qlib}), the agent summarizes relevant GitHub repository topics from the query and sequentially searches these topics to identify the most suitable repository.

\begin{table}[!t]
    \centering
    \resizebox{\linewidth}{!}{
    \begin{tabular}{cccc}
    \toprule
    \makecell{} & \textbf{\makecell{Application\\Easy}} & \textbf{\makecell{Application\\Medium}}  & \textbf{\makecell{Application\\Hard}} \\ 
    \midrule
    \textbf{\makecell{Environment\\Easy}} & \makecell{Pyflowchart\\Bolt\\yolov5} & \makecell{OCRmyPDF\\Rembg} & \makecell{TenCirChem\\ChemFormula\\Chemlib}\\
    \midrule
    \textbf{\makecell{Environment\\Medium}} & \makecell{MultiQC\\Photon\\Smap} & \makecell{Bandit\\recognize-everything} & \makecell{Aizynthfinder\\mermaid-cli}\\
    \midrule
    \textbf{\makecell{Environment\\Hard}} & Latex-OCR & \makecell{Bring-Old-Photos-\\Back-to-Life} & \makecell{qlib\\ PlotNeuralNet}\\
    \bottomrule
    \end{tabular}
    }
    \caption{GitHub repositories classified by $9$ types of difficulties.}
    \label{tab:category}
\end{table}

\begin{figure*}[!t]
\centering
\includegraphics[width=0.9\linewidth]{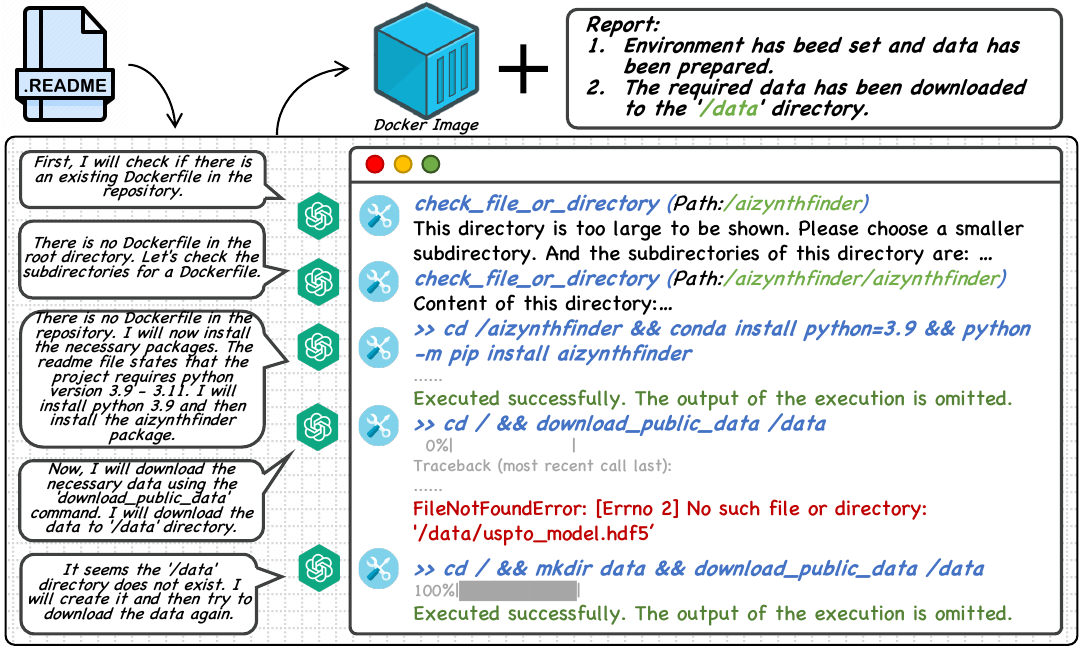}
\caption{An example about the repository setup of \texttt{AiZynthFinder}.}
\label{fig:case_setup_aizynthfinder}
\end{figure*}

\begin{figure*}[!t]
\centering
\includegraphics[width=\linewidth]{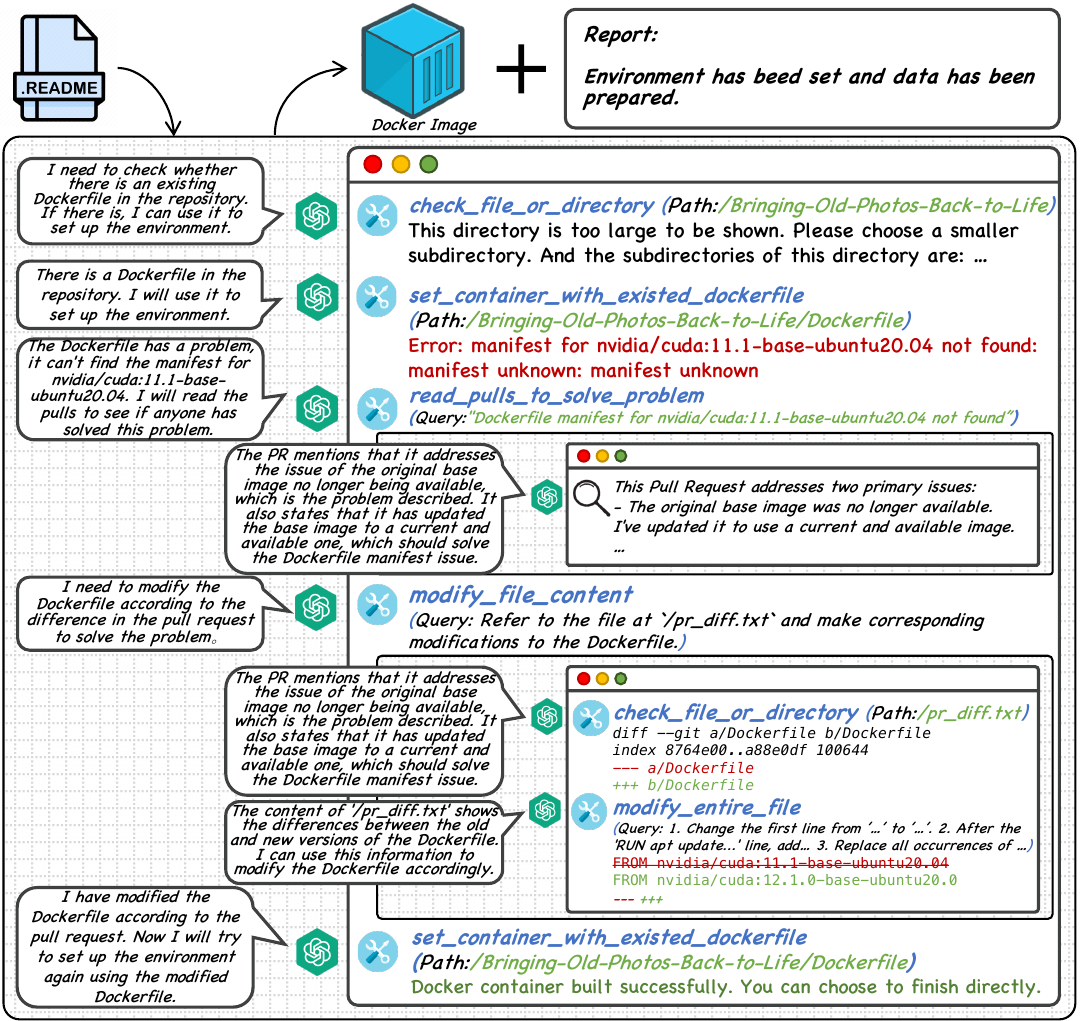}
\caption{An example of Environment Setup~(\texttt{Bringing-Old-Photos-Back-to-Life}).}
\label{fig:case_setup_bringing}
\end{figure*}

\textbf{Dynamic Handling of Setup Challenges.}
The agent is proficient in managing setup processes, even in the presence of bugs or incomplete information in the official repository documentation. 
For repositories like \texttt{AiZynthFinder}, with comprehensive setup instructions in the README (Figure~\ref{fig:case_setup_aizynthfinder}), \model efficiently follows the guidelines to set up the environment.
Conversely, for repositories such as \texttt{Bringing-Old-Photos-Back-to-Life}, although it provides an official dockerfile to build the execution environment, there exist some bugs in it.
In this situation, \model proactively seeks solutions by reviewing pull requests (PRs) and modifies the dockerfile accordingly to rectify the bugs (Figure~\ref{fig:case_setup_bringing}).

\begin{figure*}[!t]
\centering
\includegraphics[width=0.9\linewidth]{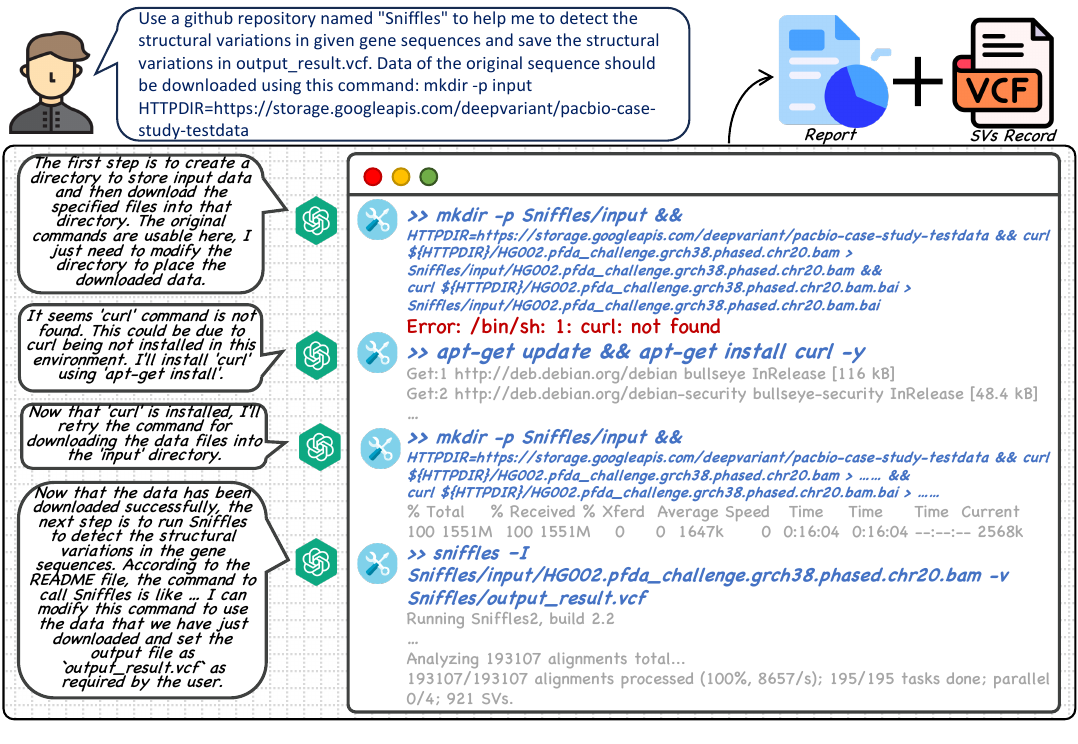}
\caption{An example about the repository apply of \texttt{Sniffles}.}
\label{fig:case_apply_sniffles}
\end{figure*}

\textbf{Efficient Execution of Complex User Queries.}
\model capably handles the execution phase, addressing various challenges that arise with complex user queries. 
An example includes a scenario where the agent needed to install additional tools (e.g., ``curl'') mid-process to fulfill a data download request (Figure~\ref{fig:case_apply_sniffles}). 
In another instance, for a query requiring transaction details in Qlib (Figure~\ref{fig:case_apply_qlib}), \model resolved a missing directory issue by consulting GitHub Issues and adapting the configuration file to generate the required output.
For example, a query asks \model to download specific data for analysis (see in Figure~\ref{fig:case_apply_sniffles}).
During the apply process, the agent found that the ``curl'' had not been installed and then it used ``apt-get'' to install ``curl'' before going on.
For another example, a user query asks \model to give the transaction details (see in Figure~\ref{fig:case_apply_qlib}).
\model first ran the repo but when it tried to get the transaction details in the ``record'' directory, it found that there exists no ``record'' directory in the repository.
Next, \model resorted to the Issue from GitHub and found an issue that can solve the record output problem.
Then, it modified the configuration file according to the issue content and re-run the command.

All the above phenomena demonstrate the robustness and flexibility of our method which can handle various non-standardized GitHub repositories to extend them as tools to accomplish user queries.
The agent effectively navigates and utilizes non-standardized GitHub repositories, extending their functionalities to meet diverse user queries. 
This success can be attributed to our designed human experience learning, which enables \model to focus on resolving subtasks by referring to human practice experience. 

%% file: section/prompts.tex
\lstset{
    numbers=none,
    keywordstyle= \color{ blue!70},
    commentstyle= \color{red!50!green!50!blue!50},
    frame=none,
    rulesepcolor= \color{ red!20!green!20!blue!20} ,
    framexleftmargin=2em,
    columns=fullflexible,
    breaklines=true,
    breakindent=0pt,
    basicstyle=\ttfamily
}

\subsection{Main Agent}

\begin{tcolorbox}[left=0mm,right=0mm,top=0mm,bottom=0mm,boxsep=1mm,arc=0mm,boxrule=0pt, frame empty, breakable]     \small     \begin{lstlisting}
You are a professional programmer. Given a query, your task is to search for a github repository and use it to solve the query.

You should make sure the result of `apply` function well completed the query. If it is lack of required elements, you can call `apply` again if you think the result is close to what you want and you think this repository can be used to solve your query. You can also call `search_by_query` function to find another repository if you think this repository is not suitable for your query.
\end{lstlisting} \end{tcolorbox}

\subsection{Search Agent}
\begin{tcolorbox}[left=0mm,right=0mm,top=0mm,bottom=0mm,boxsep=1mm,arc=0mm,boxrule=0pt, frame empty, breakable]     \small     \begin{lstlisting}
You are a professional programmer. Given a task, you want to find a github repository to solve the task.
\end{lstlisting} \end{tcolorbox}

\begin{tcolorbox}[left=0mm,right=0mm,top=0mm,bottom=0mm,boxsep=1mm,arc=0mm,boxrule=0pt, frame empty, breakable]     \small     \begin{lstlisting}
You are a professional programmer. Given a task, you want to find a github repository to solve the task. Now, your colleagues have explored some repositories. If you think any of the repository(s) might can solve your task, call `use_existing_repository` function to use it. Otherwise, call `find_a_new_repository` function to find another repository.

You will be given the query of the task and name(s) and description(s) of existed repositories.

Repository's name: {{name of repository 1}}
Description: {{description  of repository 1}}

Repository's name: {{name of repository 2}}
Description: {{description  of repository 2}}
......
\end{lstlisting} \end{tcolorbox}

\subsection{Setup Agent}
\begin{tcolorbox}[left=0mm,right=0mm,top=0mm,bottom=0mm,boxsep=1mm,arc=0mm,boxrule=0pt, frame empty, breakable]     \small     \begin{lstlisting}
You are a professional programmer. Your task is to set up the environment of the repository and prepare necessary data.

You will be provided with the readme file of the repository. You can also use `check_file_or_directory` function to check the `/<==repo_name==>` directory whether there is a existed Dockerfile. If setting up the environment is complex and there is an existing dockerfile, you can use `set_container_with_existed_dockerfile` function to directly use that dockerfile. If there is any problem with the dockerfile, you can try to use `read_pulls_to_solve_problem` function to see the pulls of this repository to solve the problem. However, `read_pulls_to_solve_problem` should not be used for reasons other than troubleshooting issues with the Dockerfile. If the existed dockerfile is built successfully, you can call `submit` function directly with property "work_directory" marked because the required docker container has already been built.

Usually the dockerfile is close to `/<==repo_name==>`, so if you don't find it in one or two try, it means there isn't a dockerfile in this repository. You don't need to try more times.

If there is no existing dockerfile, you should analyze the readme file and derive the necessary commands and execute them to set up the environment of the repository and prepare necessary data in a given container, whose base image is 'continuumio/miniconda3'. If error happens due to inappropriate base image, you can use `echo` to create a dockerfile yourself, with proper base image and necessary packages, and build it.

While operating, please note the following points:
- The commands will be run in a docker container. You don't need to use virtual environments, use the base environment only. Use pip or conda to install packages. In special cases, you can use apt-get to install necessary packages. If you use apt-get, do not forget to use apt-get update and --fix-missing.
- Any command requiring execution in a specific directory should be reformulated as: `/bin/sh -c "cd <specific directory> && <commands to be executed in this directory>"`. Every command must start with '/bin/sh -c " cd ' to locate a specific directory.
- The repository have been clone to the root directory at `/<==repo_name==>`.
- Follow the sequence of the commands, install all necessary packages first.
- Never create or activate any conda environment even if the readme requires or does so. You should install the packages in the base environment.
- If you have problem with the version of python, please reinstall python of the appropriate version with `conda install python=<version>`.
- If a function you called return you with a file path, you should pass the file path to the next function you call if need.
- If there are different choices to do the same task and you failed using one of them, you can try another alternative.

Your commands should be the parameter of the `execute_command` function. Each time you should send one or many commands. The `execute_command` function will run the commands and return the output of the commands.

In this step, you should just set up the environment and prepare the data. You don't need to run other programs or train the model.
\end{lstlisting} \end{tcolorbox}

\subsection{Apply Agent}
\begin{tcolorbox}[left=0mm,right=0mm,top=0mm,bottom=0mm,boxsep=1mm,arc=0mm,boxrule=0pt, frame empty, breakable]     \small     \begin{lstlisting}
You are a professional programmer. Your task is to utilize a github repository to solve a given query. You will operate in a docker container.

Note that it has been ensured that the repository's environment has been set up and all the data required by the readme has been fully prepared, so you mustn't execute any command to set up the environment or prepare the data or check relevant files about the environment or data anymore, unless the user provide you with a link to download necessary data. <==data_path==>

Also, all the dependencies have been installed in the base environment, please don't switch to any other conda environment. If you find you lack of any packages or tools while operating, use pip, conda or apt-get to install it. If you use apt-get, do not forget to use `apt-get update` and `--fix-missing`.

Your goal is to study the readme file especially the command lines in it and call appropriate functions to utilize the repository to solve the query. Do not execute any command to get result that you can't perceive yourself, like starting a server.

Note that the default configuration of the final executable file may not meet the demand of the query. If there is any special demands in the query, you should check the final executable file to check whether it meets the demand of the query. If not, you should make proper modification(s).

If you run a command and find the result lack of required element(s), which may because the repository itself doesn't support relevant function, you can check the issues to try to solve the problem.

If you need to deal with files provided by the user, you should firstly use `upload_directory_to_container` to upload it from local to the docker container. By default, the path claimed in the query is local path, you need to upload it. If required message can be retrieved from the output of execution of the program, summarize it to natural language and submit it. If any file is generated to answer the query, you should use `download_directory_from_container` to download the file from the docker container to local before you submit if necessary. You should also ensure required directories all exist before running a program.

We only have CPU. If the repository doesn't ask for configuration of device, ignore it.

Readme:{{readme}}
\end{lstlisting} \end{tcolorbox}

\subsection{Modify Agent}
\begin{tcolorbox}[left=0mm,right=0mm,top=0mm,bottom=0mm,boxsep=1mm,arc=0mm,boxrule=0pt, frame empty, breakable]     \small     \begin{lstlisting}
You are a professional programmer. Your task is to make modification(s) to code files to meet the given requirement. You will be given the query of modification, the content of a file and the path to the file. If you think you can meet the query through modifying this file, you can modify this file.

If the query contains path that contains information for modification, transmit that path at "query_file_path" in "modify_entire_file". You don't need to check the query file yourself, because you may neglect important message by checking and summarizing, just pass the query path and let "modify_entire_file" function to decide.

Code relevant to the query may not always reside in the currently provided file. In such cases, you should analyze the `from...import...` or `<module name>...` sections to suggest potential target file paths.

If the target path in the current file is relative path, you should decide the target file based on the current file's path.

If it starts from a module's name, which suggests the file is a python package, the file is in `/opt/conda/lib/python3.11/site-packages/<package name>` directory (python version should be decided by using `which pip`). Don't forget the suffix of the file.

You might need to locate the target file by checking the content of the files recursively. After the target file is located, you should use proper functions to modify the code.
\end{lstlisting} \end{tcolorbox}

\subsection{Judge Agent}
\begin{tcolorbox}[left=0mm,right=0mm,top=0mm,bottom=0mm,boxsep=1mm,arc=0mm,boxrule=0pt, frame empty, breakable]     \small     \begin{lstlisting}
You are a professional programmer. Your task is to judge how good a programmer use a github repository to handle a query. You will be given query and the actions the programmer took to handle the query. If the task includes input or output file, you will be given path to programmer's output. Path to input is in the query and path to the ground truth outcome will be given if there is ground truth. You can check the content in these paths and use proper ways to judge the relevance of different files. If the files are readable you can directly check them. If not, you can use the provided functions to check the md5 hash value of the files or compare the similarities of different images. Note that you can only check directory or file saved in local. If no input path, output path, truth path is given, do not check file or directory, just score based on the log.

// For ReAct & ReAct + Summary
The rule of scoring is as follows. The initial score is 0. You will be given the log of user calling functions to use the repository. For correctly setting up the environment and preparing the data, 2 point should be added for environment and 1 point should be added for data. If no data is required, point for data should be added.\nIn the given application phase, 0~4 scores should be added based on the performance. You should judge the performance based on whether it follows the instruction in the readme. If right actions(including commands and function calling) are taken and get a result, you should add 4. If asked configuration is not applied or wrong actions are taken, minus 1 point for each fault based on 4. If ground truth is provided, if the result of the application is not correct, minus 1 point.\nIn conclusion, the final score is the sum of the scores of the setup (0~3) and application phase (0~4).

// For GitAgent
The rule of scoring is as follows. The initial score is 0. You will be given the log of user calling functions to use the repository, without the steps the environment is setup.\nIn the given application phase, 0~4 scores should be added based on the performance. You should judge the performance based on whether it follows the instruction in the readme. If right actions(including commands and function calling) are taken and get a result, you should add 4. If asked configuration is not applied or wrong actions are taken, minus 1 point for each fault based on 4. If ground truth is provided, if the result of the application is not correct, minus 1 point.\nGenerally, if valid output is given, the score should be 4.

Query:{{query}}

Action:{{action_log}}

Input path:{{input_path}}

Output path:{{output_path}}

Ground Truth path:{{truth_path}}
\end{lstlisting} \end{tcolorbox}